\documentclass[twocolumn,fleqn,titlepage]{aa}
\newcommand{\pd}[2]{\frac{\partial #1}{\partial #2}}
\usepackage{graphicx}

\begin{document}


\title{Oscillations of tori in the pseudo-Newtonian potential}
\author{Eva \v Sr\'amkov\'a \inst{1} \and Ulf Torkelsson \inst{2} \and
Marek A. Abramowicz \inst{2,3}}

\institute{Institute of Physics, Silesian University in Opava, Bezru\v covo
n\'am. 13, 746 01 Opava, Czech Republic \and
Department of Physics, G\"oteborg University, 412 96 G\"oteborg, Sweden
\and Copernicus Astronomical Centre, Bartycka 18, PL-00-716 Warszawa, Poland}

\date{Received / Accepted}


\abstract
{The high-frequency quasi-periodic oscillations (HF QPOs) in neutron star and 
stellar-mass black hole X-ray binaries may be the result of a resonance between 
the radial and vertical epicyclic oscillations in strong gravity.}
{In this paper we investigate the resonant coupling
between the epicyclic modes in a torus in a strong gravitational field.}
{We perform numerical simulations of axisymmetric
constant angular momentum tori in
the pseudo-Newtonian potential.  The epicyclic motion is excited by adding
a constant radial velocity to the torus.}
{We verify that slender tori perform epicyclic motions at the
frequencies of free particles, but the epicyclic frequencies decrease as
the tori grow thicker.  More importantly, and in contrast to previous
numerical studies, we do not find a coupling between the radial and
vertical epicyclic motions.
The appearance of other
modes than the radial epicyclic motion in our simulations is rather due to
small numerical
deviations from exact equilibrium in the initial state of our torus.}
{We find that there is no pressure coupling between the two 
axisymmetric epicyclic modes 
as long as the torus is symmetric with respect to the equatorial plane.
However we also find that there are other modes in the disc that may be more
attractive for explaining the HF QPOs.}


\maketitle


\section{Introduction}

Observations by Rossi-XTE have revealed that some accreting neutron stars (NS)
and stellar-mass black holes (BH) show quasi-periodic oscillations
in their X-ray fluxes with frequencies of 100 - 1\,300 Hz (see van der
Klis \cite{klis00} for a review).  Some X-ray sources are 
simultaneously varying on two frequencies that form a ratio of 3:2. For BH 
QPOs this was first noticed by Abramowicz \& Klu\'zniak (\cite{AK01}) 
(see also Strohmayer \cite{stro01},
McClintock \& Remillard \cite{mc:rem}). Abramowicz et
al. (\cite{abr03}) argued that this is also the case for NS QPOs: although 
the ratio is varying, it clearly distinguishes the same 3:2 value (see Belloni 
et al. \cite{bel05} for a criticism).  
The mechanism that is responsible for this phenomenon is not fully understood. 
It has been
pointed out by Klu\'zniak \& Abramowicz (\cite{klu00}) that the rational frequency
ratio could be formed naturally in a resonance between two modes of accretion disc 
oscillations. Radial and vertical epicyclic oscillations in an accretion disc have 
been the most often discussed possibility (e.g. Abramowicz \& 
Klu\'zniak \cite{AK04}, or a collection of review articles in Abramowicz 
\cite{abr05}).
In Newtonian gravity with the $1/r$ potential the epicyclic
frequencies $\omega_R$ and $\omega_z$ are both equal to the
Keplerian frequency $\Omega_{\rm K}$. This is no longer the case 
in the strong gravitational field close to a compact
object, where $\omega_z > \omega_R$, which allows the two frequencies to
assume a 3:2 commensurability at some location. 

Several authors have numerically studied oscillations of thick accretion discs
around black holes and neutron stars in the QPO context in both the Newtonian 
(Rubio-Herrera \& Lee \cite{rub05}a,b) and the relativistic regime
(Rezzolla et al. \cite{rez03}a,b, Montero et al. \cite{mon04}, Zanotti et 
al. \cite{zan05}). 
Recently, Lee, Abramowicz \& Klu\'zniak (\cite{Lee04}) have performed numerical
simulations of pseudo-Newtonian slender tori in order to study the response
of the torus to impulsive and periodic perturbations associated with the
spin of the neutron star.
They reported that a vertical epicyclic motion was excited
in the torus due to the non-linear coupling between the epicyclic modes when the 
applied
perturbation was purely radial, and found its response to a periodic 
perturbation to be dependent
on the difference between the spin frequency and the frequency difference 
between the two epicyclic 
modes 
(see also Klu\'zniak et al. \cite{klu04} and Klu\'zniak, Abramowicz \& Lee
\cite{klu042}).

In this paper we present a numerical study of oscillatory modes
of slender tori in equilibrium that are given a small impulsive radial perturbation.   
We start in Sect. 2 by writing down the hydrodynamic equations, from which
we construct our equilibrium tori, and then we describe the numerical 
method that we use to follow the time evolution of the tori.
The results
of our numerical simulations are described in Sect. 3, while
Sect. 4 is devoted to a discussion of the results and our conclusions.


\section{Model}

\subsection{Hydrodynamics}

We use the equations of
nonrelativistic ideal hydrodynamics to describe the oscillations of a torus
around a black hole. 
Ignoring the effects of radiative transport and self-gravity, the equations 
take the form
\begin{eqnarray}
\pd{\rho}{t} + {\bf \nabla} \cdot (\rho {\bf v}) & = & 0, \label{cont_eq}  \\
\rho \pd{{\bf v}}{t} + (\rho {\bf v} \cdot {\bf \nabla}) {\bf v} &  =  &
-\nabla P -\rho \nabla \Phi, \label{euler_eq}\\
\pd{(\rho\epsilon)}{t} + \nabla \cdot (\rho \epsilon {\bf v}) & = &
-P {\bf \nabla} \cdot {\bf v}, \label{en_eq}
\end{eqnarray}
where $\rho$ and $\epsilon$ are the density of mass and the specific internal 
energy, 
${\bf v}$ is the fluid velocity, $\Phi$ the gravitational potential and $P$ is 
the pressure, which is given by the equation of state for an ideal gas, 
$P = \rho\epsilon (\gamma-1)$, with $\gamma = 4/3$. 

\begin{figure}
\resizebox{\hsize}{!}{\includegraphics{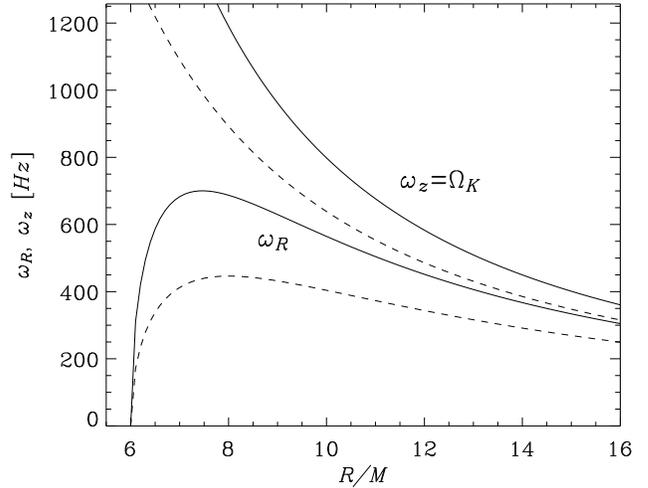}}
\caption{Radial ($\omega_R$) and vertical 
($\omega_z=\Omega_{\rm K}$) epicyclic frequencies
in the Paczy\'nski \& 
Wiita potential, and the corresponding 
radial (lower dashed curve) and vertical (upper dashed curve) epicyclic frequencies
in the Schwarzschild metric, for a $M=10$ $M_{\odot}$ black hole.}
\label{PN}
\end{figure}

The most important general relativistic effects in the Schwarzschild metric 
are mimicked by using the pseudo-Newtonian potential 
(Paczy\'nski \& Wiita \cite{paczynski:wiita})  
\begin{equation}
\Phi = -\frac{GM}{(r-r_{\rm g})}, 
\end{equation}
where $r$ is the spherical radius and $r_{\rm g} = 2GM/c^2$ is the 
Schwarzschild radius of the black hole. In this potential, the epicyclic 
frequencies have qualitatively the same behaviour as in the Schwarzschild 
spacetime, i.e. $\omega_R<\omega_z=\Omega_{\rm K}$ (see Fig. \ref{PN}), and are 
explicitly given by $\omega_R= 2\pi \nu_R= \sqrt{GM(R-3r_g)}/\sqrt{R(R-r_g)^3}$ and 
$\omega_z=2\pi \nu_z=  \sqrt{GM/R}/(R-r_g)$. 
In the Schwarzschild metric, the frequencies assume the 3:2 ratio at $R=10.8$ $GM/c^2$, 
while in the Paczy\'nski \& Wiita potential this resonance occurs at $R=9.2$ $GM/c^2$.

\begin{figure}
\resizebox{\hsize}{!}{\includegraphics{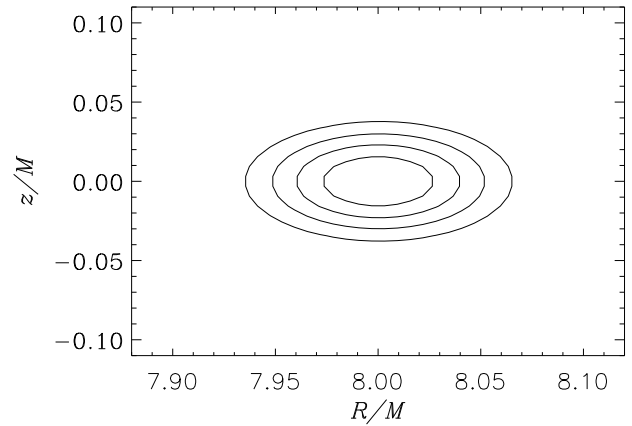}}
\caption{Density contours in a meridional cross-section of a 
slender torus at $R=8$ $M$
with radial extent of approximately $0.15$ $M$.}
\label{torus}
\end{figure}

We construct a non-accreting torus in hydrostatic equilibrium with constant
specific angular momentum 
$l = l_0$ as our initial state.  For convenience we will adopt geometric units,
where $G = c = 1$, for the rest of the paper.
The only non-zero velocity component is then
the azimuthal component, which
in cylindrical coordinates $\{R,\phi,z\}$ is given by $v_\phi = \Omega R$. The
equilibrium configuration is then governed by the time-independent Euler 
equation 
\begin{equation}
-\frac{1}{\rho}\nabla P = \nabla \Phi - \Omega^2 {\bf R} .
\end{equation}
For a barotropic fluid, in which $\Omega = \Omega (R)$, we can define 
a rotational potential 
$\Phi_{\rm rot} = \int \Omega^2 R  {\rm d}R = \int l_0^2/R^3  {\rm d}R$.
In addition we use a polytropic equation of state,
$P = K\rho ^\gamma $, where $K$ denotes the polytropic 
constant.  The Euler equation can then be integrated to give an expression for 
the equipotentials $W$ that determine the shape of the torus: 
\begin{equation}
\frac{\gamma}{(\gamma-1)} \frac{P} {\rho} + \Phi_{\rm eff} + C = W = \mbox{const}.
\end{equation}
Here $\Phi_{\rm eff} = \Phi + \Phi_{\rm rot}$ is the effective potential and $C$ an 
integration constant defining the surface of the torus. 
For particular values of the specific angular momentum, an equilibrium torus 
can 
now be constructed by filling the appropriate equipotential surfaces with 
a rotating fluid. An example of the meridional 
cross-section of a slender (in the sense 
that its radial extent is much smaller than its distance from the central 
object) torus is shown in Fig. \ref{torus}.

\subsection{Numerical simulations}

The numerical simulations in this paper are performed using the 
two-dimensional ZEUS-2D code by Stone \& Norman (\cite{sto92}) that solves Eqs. 
(\ref{cont_eq})-(\ref{en_eq}). 
ZEUS is a time-explicit, 
Eulerian finite difference code that uses a staggered grid. Shock waves are 
handled by an artificial viscosity tensor.
The magnetic field, 
self-gravity and radiation transport in ZEUS were switched off during our
simulations. 

We study two different groups of models
(Tab. \ref{models} ). Models 
 A1-A4 represent slender tori with a radial extent of approximately $0.15$ $M$ at 
different radial positions (see Fig. \ref{torus}).
Models B1-B4 make up a series of thicker (non-slender) 
tori. We perturb the tori by adding a velocity field that is constant in space 
at 
$t=0$. 
The perturbation is always subsonic, corresponding to a Mach number at the torus 
center of mass of 0.3.

\begin{figure}[t!]
\resizebox{\hsize}{!}{\includegraphics{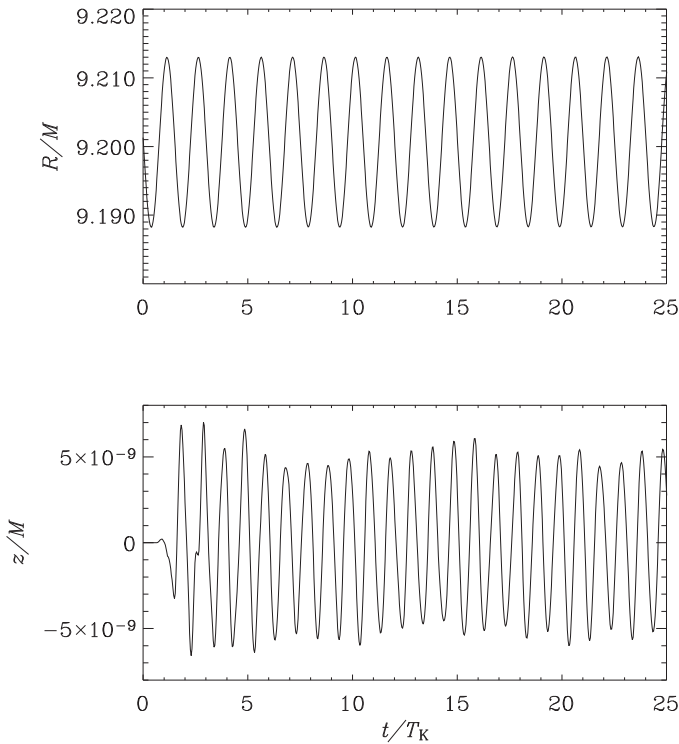}}
\caption{Radial (top) and vertical (bottom) oscillations of a torus centered at 
$R=9.2$ $M$ (Model A3) where the two epicyclic modes are in a 3:2 ratio. The time 
is given in units of the Keplerian period $T_{\rm{K}}$ defined at the centre 
of the torus.
The torus performs a radial oscillation with an amplitude of 0.01 $M$. The vertical 
oscillation excited here is seven orders of magnitude smaller than the radial 
oscillation.} 
\label{8M}
\end{figure}

We use a uniform grid in cylindrical coordinates in all our simulations and 
the resolutions of the different models are specified in Tab. \ref{models}.

\begin{figure}[t!]
\resizebox{\hsize}{!}{\includegraphics{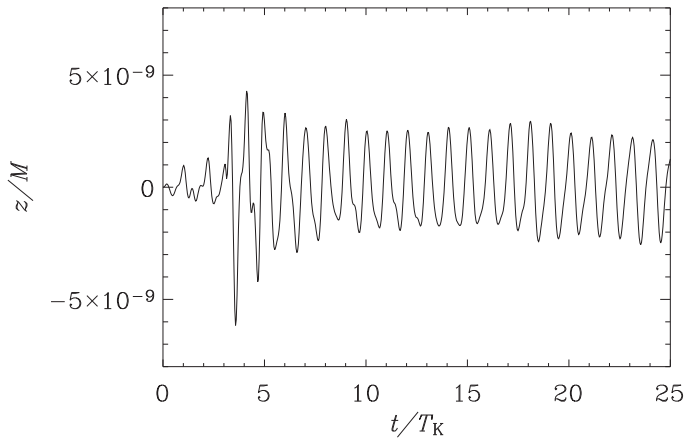}}
\caption{ Same as the bottom panel in Fig. \ref{8M}, but using a finer grid 
(Model A4). Note the decrease of the amplitude of the oscillation compared to Fig. 
\ref{8M}.}
\label{8Mresol}
\end{figure}

\begin{figure}[t!]
\resizebox{\hsize}{!}{\includegraphics{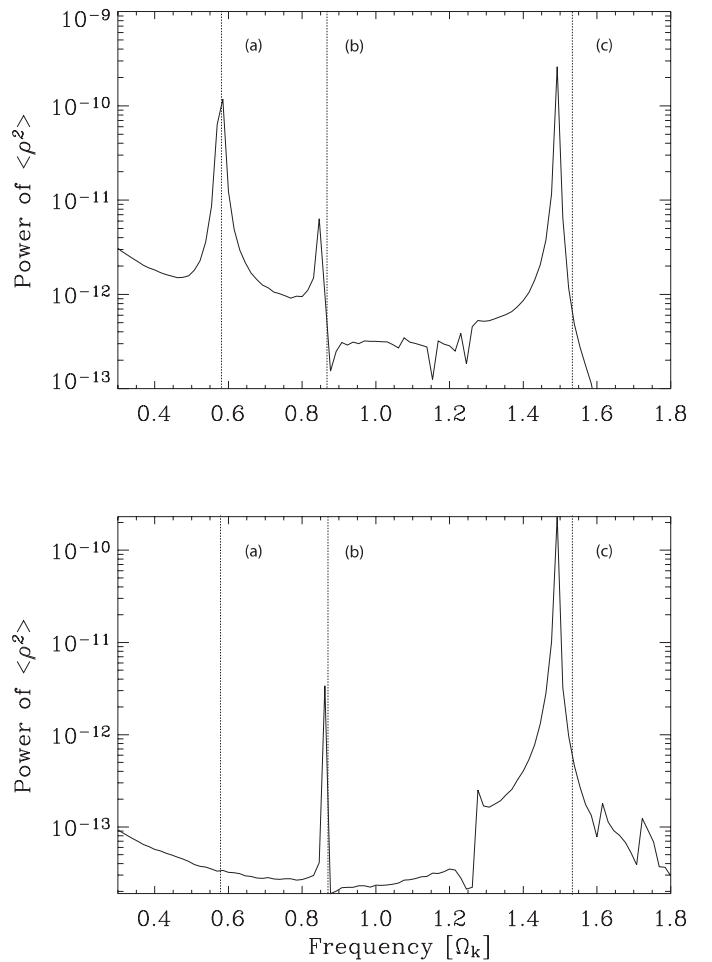}}
\caption{{\it Top}: Power spectrum of $\langle \rho^2\rangle$ of a 
radially 
perturbed slender torus (Model A2). Frequencies are given in units of 
$\Omega_{\rm K}$, the Keplerian angular velocity at the centre of the
torus. The left peak at approximately 0.58 
$\Omega_{\rm K}$
corresponds to the radial epicyclic motion (see Tab. \ref{models}). 
The other main peaks are located at about
0.86 $\Omega_{\rm K}$ and 1.5 $\Omega_{\rm K}$. The vertical lines denote the 
frequencies of the radial epicyclic mode (a), the plus-mode (b) and 
the breathing mode (c). 
 {\it Bottom}:  Same as the top panel, but for an unperturbed 
torus. The peak  belonging to the radial motion is absent, but the other two peaks 
remain with approximately the same power, as in the top panel.}
\label{powerD}
\end{figure}

\begin{figure}[t!]
\resizebox{\hsize}{!}{\includegraphics{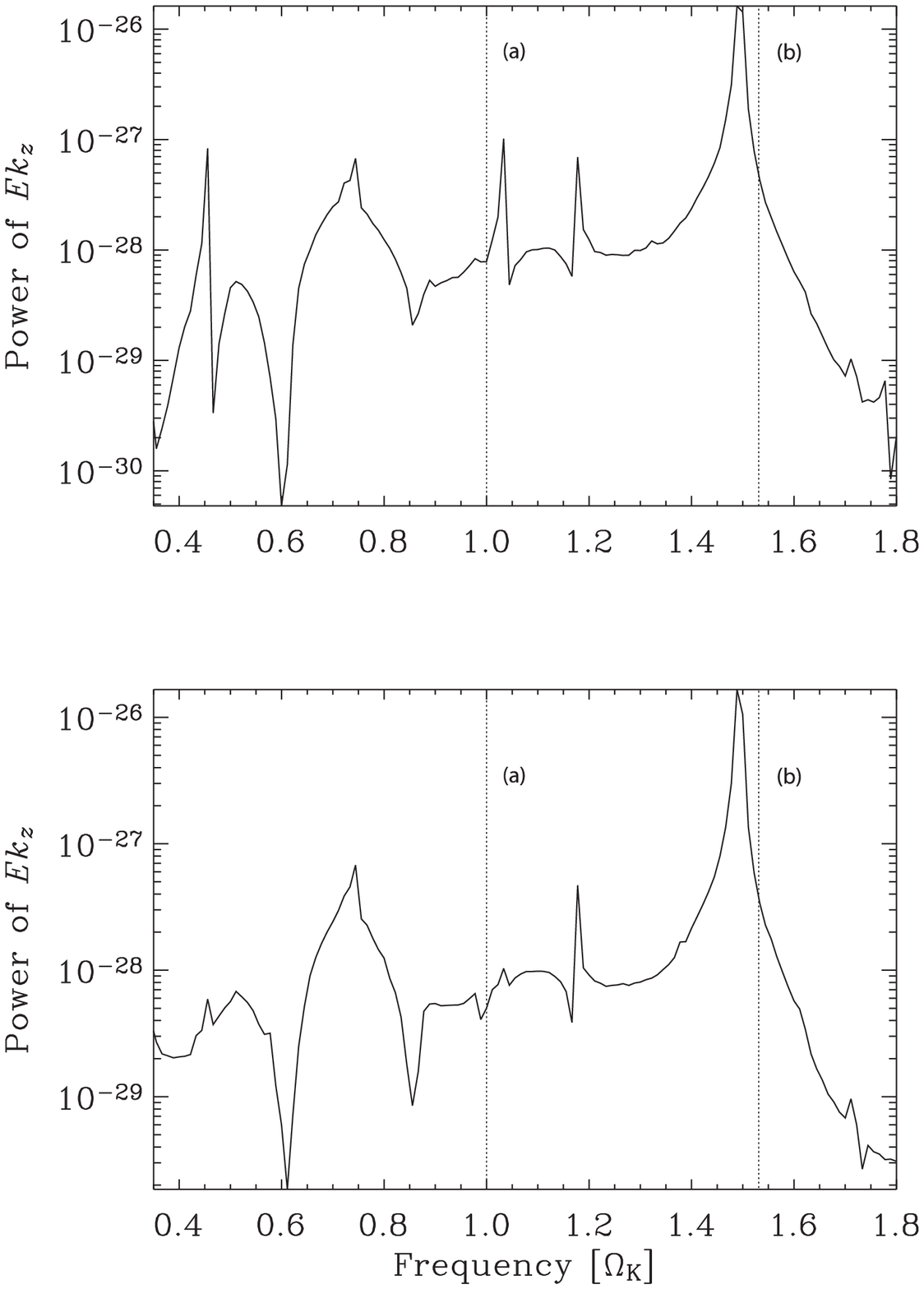}}
\caption{{\it Top}: Power spectrum of the total vertical kinetic energy 
of  a 
radially perturbed slender torus centered at $R = 8$ $M$ (Model A2). Frequencies 
are given in units of $\Omega_{\rm K}$. Note the major peak at 1.5 
$\Omega_{\rm K}$  and the smaller peaks at 0.45 $\Omega_{\rm K}$, 0.75 
$\Omega_{\rm K}$, 1.05 $\Omega_{\rm K}$, and 1.2 $\Omega_{\rm K}$. The 
vertical lines denote the 
frequencies of the vertical epicyclic mode (a) and the breathing mode (b) 
calculated for this Model.
{\it Bottom}: Same as the top panel, but for an unperturbed torus. The 0.45 
$\Omega_{\rm K}$ and 1.05 $\Omega_{\rm K}$ peaks are missing in this case}.
\label{powerE}
\end{figure}


\section{Results}

For the data analysis we look at some of the characteristic quantities as
functions of time, 
such as the total kinetic energy, the mean of 
the square of the density (as was previously done by e.g. Zanotti et al. \cite{zan05}  
and Blaes et al. \cite{bla06a}a),
and (following Lee et 
al. \cite{Lee04}) the radial and vertical positions of the centre of mass of the torus,
as well as snapshots of the density, internal energy and velocity 
field inside the torus. 

\begin{figure*}[t!]
\centering
\resizebox{\hsize}{!}{\includegraphics{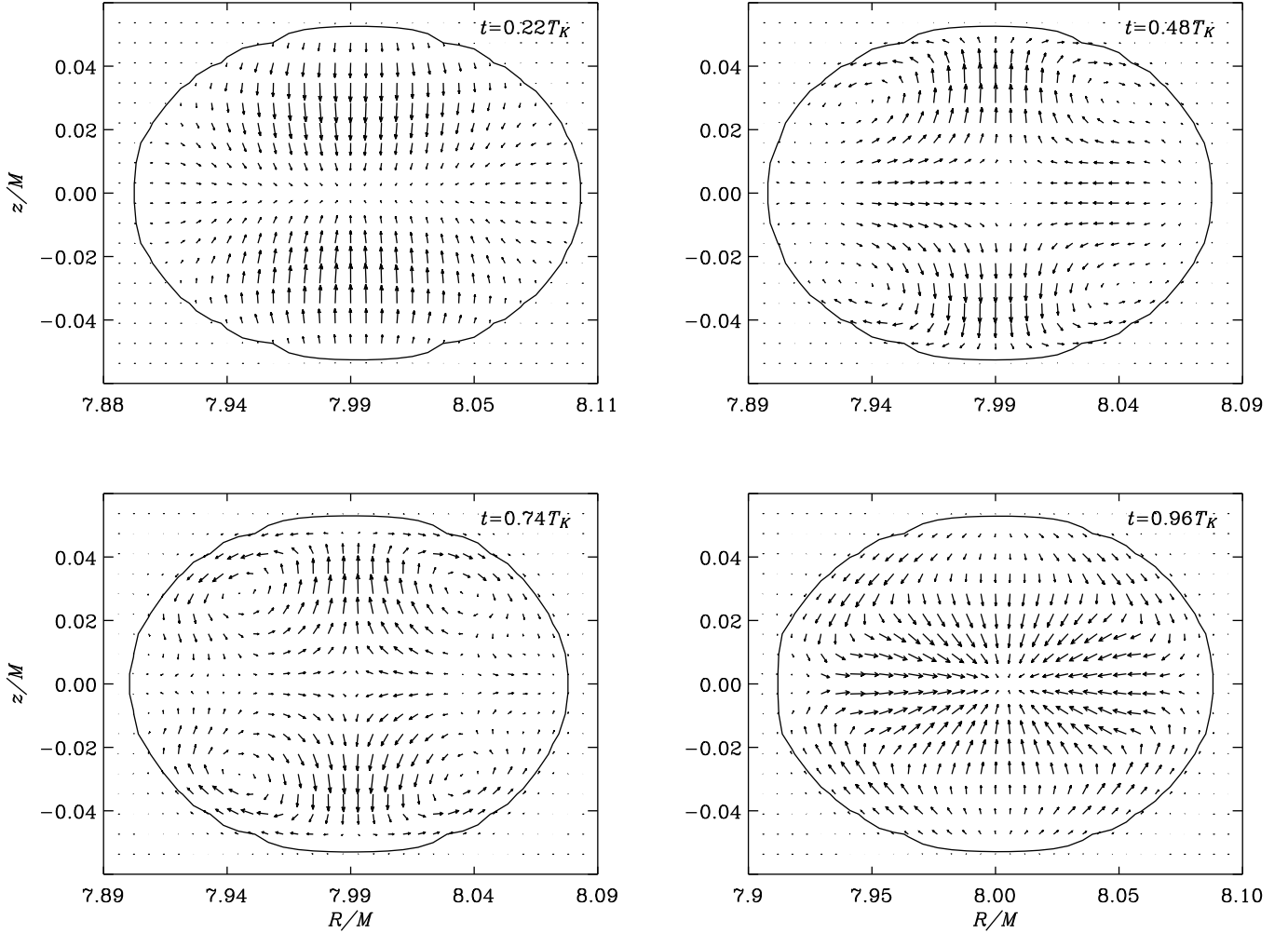}}
\caption{Snapshots of the $\rho {\bf v}$ inside the torus (Model A2) 
after substraction of the radial epicyclic motion. The snapshots cover
a period of the breathing mode, $T=(1/1.54)$ $T_{\rm K}$, where $T_{\rm K}$ 
again denotes the Keplerian period at the centre of the torus. 
The residual $\rho {\bf v}$ resembles that
of the breathing mode (see Fig. 1 in Blaes et al. \cite{bla06a}a).}
\label{velovect}
\end{figure*}

\subsection{Slender torus oscillations}

We consider a slender torus (Models A1-A4) and perturb it slightly 
by adding a purely radial constant velocity field at $t=0$. The torus then 
performs radial oscillations at the radial epicyclic frequency of a test particle
in a circular orbit at the centre of the torus  (Fig. \ref{8M}). 
We also find a
small oscillation in the vertical direction at the vertical 
epicyclic frequency of a test particle (see Tab. \ref{models} for the 
frequencies measured from the simulations), but the amplitude of this
motion is more than six orders of magnitude smaller than that of the 
radial epicyclic motion.  The small amplitude of the vertical oscillation is
independent of both the amplitude of the radial epicyclic motion and the
radial position of the centre of the torus. Moreover, it 
decreases with increasing numerical resolution, as shown in Fig. \ref{8Mresol}.  
Our conclusion is therefore that it is not the result of a resonance between 
the two epicyclic modes.
Rather it is excited by numerical noise in the simulations. 

The power spectra of the mean of the square of the density and the total 
vertical kinetic energy 
for Model A2 can be seen in  Figs. \ref{powerD} 
and \ref{powerE}. 
The $\langle \rho^2\rangle$ power spectrum is not sensitive to incompressible modes, 
such as the 
vertical epicyclic mode, but the radial epicyclic mode is weakly compressible because 
of the cylindrical geometry, and thus we can see it in Fig. \ref{powerD}.
It corresponds to the left peak at $\omega_R = 0.57$ $\Omega_{\rm K}$. 
Beyond that, there are two other peaks in the power spectrum of 
$\langle \rho^2\rangle$.  These peaks appear independently of whether 
the torus has been perturbed or not (compare the top and bottom panels of 
Fig. \ref{powerD}). 
The strong peak at 1.5 $\Omega_{\rm K}$ is present in the spectrum 
at this frequency regardless of the location of the torus, while the
less prominent peak at approximately 0.86 $\Omega_{\rm K}$ slightly changes 
its position depending on the location of the torus.

Blaes et al. (\cite{bla06a}a) have calculated the lowest-order slender 
torus modes. Two of these modes, the incompressible "plus-mode" and the acoustic 
"breathing" mode have frequencies of $0.87$ $\Omega_{\rm K}$ and $1.54$ $\Omega_{\rm K}$ 
for the torus of Model A2 (see their Tab. 2 for explicit expressions). 
These frequencies match well with the peaks at 
$0.86$ $\Omega_{\rm K}$ and $1.5$ $\Omega_{\rm K}$ in Fig. \ref{powerD}.  

The $1.5$ $\Omega_{\rm K}$ peak also appears in the power spectrum of the total 
vertical kinetic energy, as seen in Fig. \ref{powerE}. We also find four minor 
peaks in the power spectrum of the vertical kinetic energy at 0.45 $\Omega_{\rm K}$, 
0.75 $\Omega_{\rm K}$, 1.05 $\Omega_{\rm K}$, and 1.2 $\Omega_{\rm K}$. 
The peaks at 0.45 $\Omega_{\rm K}$ and 1.05 $\Omega_{\rm K}$ are absent in the case
of an unperturbed torus (bottom panel of Fig. \ref{powerE}). None of the minor peaks
have any corresponding peaks in the power spectrum of $\langle \rho^2\rangle$, and
they cannot be identified with any of the slender torus modes that Blaes et al. (\cite{bla06a}a)
calculated. An intriguing feature of Fig. \ref{powerE} is the lack of power at the 
frequency of the radial epicyclic mode, which suggests that this mode has a negative 
influence on the vertical motion.

Fig. \ref{velovect} shows four snapshots of $\rho {\bf v}$ inside the torus 
(Model A2)
after substraction of
the mean radial motion of the torus. They cover one period 
of the breathing 
mode 
$2\pi/(1.54$ $\Omega_{\rm K})$. The velocity pattern is dominated by the breathing mode, 
though we do see radial flows that match the plus-mode too.

\begin{table*}[t!]
\centering 
\caption{Properties of the models. 
From left to right the columns give the name of the model, grid resolution, 
radial ($R$-grid) and vertical ($z$-grid) boundaries of the grid, the inner
($R_{\rm in}$), central ($R_{\rm c}$), and outer ($R_{\rm out}$) radii, the 
initial
velocity kick in terms of the Mach number M at the centre of the torus,
and radial $\omega_R$ and vertical $\omega_z$ epicyclic 
frequencies of the torus centre in units of $\Omega_{\rm K}$.}
\label{models}
\begin{tabular}{cccccccccc}\\ \hline \hline 
Model & Grid points in $(R,z)$ & $R$-grid & $z$-grid & $R_{\rm in}/M$ & $R_{\rm c}/M$ & 
$R_{\rm out}/M$ & M & $\omega_R/\Omega_{\rm K}$  & $\omega_z/\Omega_{\rm K}$  \\ \hline 
A1 & 128 x 128 & 6.95 - 7.65  & -0.35 - 0.35  & 7.25 & 7.33 & 7.41 & 0.3 &
0.50  & 1.00 \\ 
A2 & 128 x 128 & 7.65 - 8.35  & -0.35 - 0.35  & 7.92 & 8.00 & 8.08 & 0.3 &
0.58  & 1.00 \\
A3 & 128 x 128 & 8.80 - 9.60  & -0.35 - 0.35  & 9.12 & 9.20 & 9.28 & 0.3 &
0.67  & 1.00 \\ 
A4 & 200 x 200 & 8.80 - 9.60  & -0.35 - 0.35  & 9.12 & 9.20 & 9.28 & 0.3 &
0.67  & 1.00 \\ 
\hline
B1 & 128 x 128 & 8.80 - 09.50 & -0.40 - 0.40 & 9.100 & 9.2 & 09.302  & 0.3 &
 0.67 & 0.99 \\ 
B2 & 156 x 156 & 7.00 - 12.00 & -2.50 - 2.50 & 7.800 & 9.2 & 11.069 & 0.3 &
 0.64 & 0.93 \\
B3 & 186 x 186 & 6.00 - 15.00 & -4.50 - 4.50 & 6.999 & 9.2 & 13.998 & 0.3 &
 0.57 & 0.76 \\
B4 & 200 x 200 & 5.50 - 17.00 & -5.75 - 5.75 & 6.000 & 9.2 & 16.503 & 0.3 &
 0.50 & 0.60 \\
 \hline
\end{tabular}
\end{table*}

\begin{figure}
\resizebox{\hsize}{!}{\includegraphics{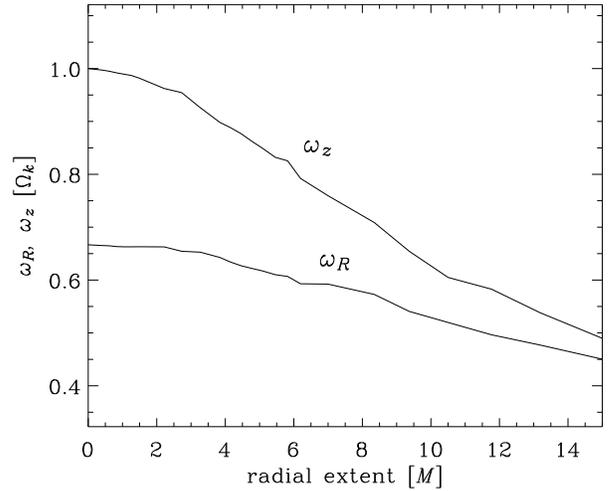}}
\caption{Radial $\omega_R$ and vertical $\omega_z$ 
epicyclic frequencies of the 
torus as a function of its width assuming that the
torus is centered at $9.2$ $M$.}
\label{shift}
\end{figure}

\subsection{Epicyclic oscillations in thick tori}

\begin{figure}[t!]
\resizebox{\hsize}{!}{\includegraphics{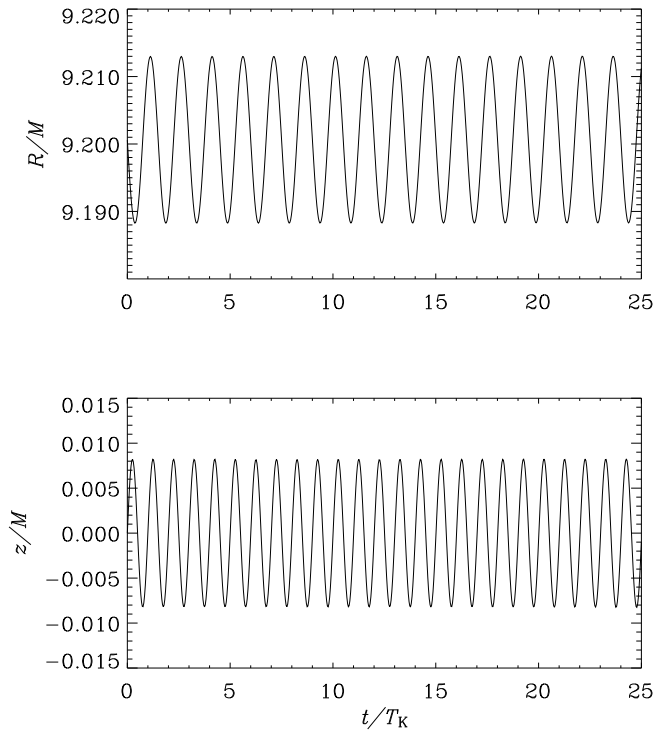}}
\caption{ Radial (top) and vertical (bottom) oscillations of a torus 
centered at $R = 9.2$ $M$ (Model A3) when both the modes are excited. Note that 
there is no exchange of energy between the two modes.} 
\label{both}
\end{figure}

In order to excite both the epicyclic modes, we applied a perturbation 
in both the radial and the vertical directions.
The resulting epicyclic frequencies for four different tori (Models B1-B4) 
are presented 
in Tab. \ref{models}. 
For an infinitely slender torus, the frequencies are consistent with the 
epicyclic frequencies of a test particle at the torus centre, 
$\omega_R$ = 0.666 $\Omega_{\rm K}$ and $\omega_z$ = $\Omega_{\rm K}$.  
The frequencies decrease as the torus grows thicker, as seen from  
Tab.\ref{models} and Fig.
\ref{shift}.
A similar trend was reported by Rubio-Herrera \& Lee  (\cite{rub05}b), and
Blaes et al. (\cite{bla06b}b), 
where it has been explored analytically for slightly non-slender Newtonian 
tori.  It is worth mentioning that even when we excite both the epicyclic 
modes, we do not see any exchange of energy between them (see Fig. 
\ref{both}).

\section{Discussion and Conclusions}

\subsection{Epicyclic oscillations}

We have performed two-dimensional numerical simulations of oscillations of 
constant angular momentum tori in the Paczy\'nski \& Wiita 
potential. The tori were 
given an impulsive radial perturbation 
and were allowed to evolve in time afterwards.  The tori are then describing 
radial epicyclic motions.  For a sufficiently slender torus the frequency of
this oscillation agrees with that of a test particle.
As the torus grows thicker we find that
the epicyclic frequencies significantly decrease.
This is in accord with previous numerical results
by Rubio-Herrera \& Lee (\cite{rub05}b), as well as recent analytic 
work where the expressions for the (negative) corrections to the epicyclic frequencies 
were derived using a perturbative method accurate to second-order in the
extent of the torus
(Blaes et al. \cite{bla06b}b).

One of the main objectives of this work is to investigate the modes that 
may be excited by a radial epicyclic oscillation in the torus. 
In particular, we look for the resonant coupling between the radial and the 
vertical epicyclic modes, which Klu\'zniak \& Abramowicz  
suggested plays a 
significant role in the high-frequency QPOs.
We find that the centre of mass of the torus
performs a vertical oscillation 
of a very low amplitude, but it is independent of the amplitude of the radial
epicyclic
oscillation and the location
of the torus, unlike what one would expect if it was due to a resonance between
the two epicyclic modes.  Rather it is likely that the vertical epicyclic
oscillation is excited by numerical noise in the simulations.
A coupling between the radial and vertical epicyclic modes in the
3:2 parametric resonance may only happen in a limited
range of the parameter space, the Arnold tongue (see e.g. Arnold \cite{arn78}, 
Arnold \cite{arn83}). 
This range is not known a priori, but one relevant parameter is the {\em initial}
vertical amplitude. 

Lee et al. (\cite{Lee04}) found that a small vertical epicyclic motion was excited 
by the radial epicyclic motion in their simulations of an axisymmetric torus. 
We believe that the reason for this is that their initial torus was not 
perfectly symmetric around the $z = 0$ plane, and thus the initial amplitude 
of the vertical oscillation was not zero.
We did not find that any vertical oscillations were excited in our simulations, 
in which the initial amplitude of the vertical motion was zero - thus in our 
case there was not any coupling between the radial and vertical epicyclic modes.
This is hardly surprising: conservation of momentum prohibits the excitation of 
a vertical epicyclic motion of the centre of the torus, if it is initially
at rest in the equatorial plane, and no external vertical forces are at play.

In order to find the Arnold tongue empirically one should perform
many simulations similar to these described in our paper, but with
different initial vertical amplitudes, and with different ratios (close
to 3:2) of the radial and vertical epicyclic frequencies. This
will give the range in the parameter space in which the mode coupling
for axisymmetric oscillations may occur.

A more promising approach might be to consider nonaxisymmetric oscillations.
The simplest example of nonaxisymmetric vertical motion is an $m = 1$ 
precessing warp in an accretion disc.
Such a warp will resonantly drive a radial
epicyclic motion (Papaloizou \& Pringle
\cite{pap83}, Papaloizou \& Lin \cite{pap95}).
A warp can be produced, for instance, 
by the radiation pressure  from a central radiation source 
(Pringle \cite{pri96}). 
The role of $m = 1$ warps for the HF QPOs has been discussed 
by Kato (\cite{kat04}, \cite{kat05}). 

Another possibility is that the two modes can be coupled by the local
fluctuations that are generated by the turbulence in the disc  (note 
that the turbulence is nonaxisymmetric).
It is today widely accepted that accretion discs become turbulent because
of the
magnetorotational instability, that was introduced by Balbus \& Hawley
(\cite{bal91}).
This turbulence generates the radial transport of angular momentum that
drives the accretion and heats the disc, but
its role in other aspects of the dynamics of accretion discs has hardly
been explored 
though.
Vio et al. (\cite{vio06}) have recently studied how
stochastic processes can produce resonances 
between the epicyclic 
modes in an almost Keplerian disc.

\subsection{Other oscillatory modes}

A problem with attributing the HF QPOs to a resonance between the two
epicyclic modes is that they only occur at a special resonance radius,
and thus require a certain amount of fine-tuning.
However our tori contain more oscillatory hydrodynamical modes, and
the power spectra of $\langle \rho^2\rangle$ reveal two significant peaks,  
both of whom occur regardless of whether there is an initial 
perturbation\footnote{These modes are excited in our simulations because 
the equilibrium torus of Sect. 2 is not any longer in perfect equilibrium when
it is implemented on our grid.}.
Firstly, there is a peak at 1.5 
$\Omega_{\rm K}$, 
which is also displayed in the power spectrum of the vertical energy.  
This peak 
appears at approximately the same frequency in all the simulations, no matter what is 
the location of the torus. 
It is of interest since 
its frequency forms a ratio of 3:2 with the Keplerian frequency
$\Omega_{\rm K}$, which is also the vertical epicyclic frequency.
It can perhaps be identified as the lowest-order acoustic wave mode 
(see Blaes et al. \cite{bla06a}a), 
which has a frequency close to 1.5 $\Omega_{\rm K}$, but with some
dependence
on the radial location of the torus. 
It was suggested by Blaes et al. (\cite{bla06a}a) that this mode and the vertical 
epicyclic mode 
might be responsible for the 3:2 commensurability observed in the HF QPOs.  

The other peak at 0.86 $\Omega_{\rm K}$ 
agrees well with the frequency
of a surface gravity wave mode that Blaes and his collaborators labeled as the 
plus-mode. This oscillatory mode was found to be excited in previous numerical 
studies with either localized (Rubio-Herrera \& Lee \cite{rub05}a,b), or global 
(Rezzolla et al. \cite{rez03}a,b, Montero et al. \cite{mon04} and Zanotti et al. 
\cite{zan05}) external perturbations, where it was attributed to one of the set 
of acoustic p-modes of oscillations. Such identification was based on numerical 
calculations of 
eigenfunctions and eigenfrequencies of vertically integrated tori, carried out 
by Rezzolla et al. 
\cite{rez03}b, and it was clarified in Blaes et al. (\cite{bla06a}a) 
(see their discussions in Sects. 4.4 and 5.1) 
that this in general incompressible mode may indeed be related to an acoustic p-mode 
in the two-dimensional case of a vertically integrated torus.
The frequencies of the p-modes were found to appear in an approximate harmonic 
sequence $2:3:4...$, in which the first frequency is consistent with the radial 
epicyclic frequency of the torus, the second is the plus-mode frequency, and others 
correspond to the higher-order p-mode frequencies. The plus-mode and the radial 
epicyclic mode 
that appear in a 3:2 ratio were proposed by Rezzolla et al. (\cite{rez03}a)
as an explanation of the black hole HF QPOs.


\acknowledgements

The authors would like to thank William H. Lee for valuable discussions on
his simulations.
Most of this work was done during the visit of E.S. to 
G\"oteborg University within the SOCRATES-ERASMUS program. We also 
gratefully acknowledge the hospitality of NORDITA in Copenhagen and Copernicus 
Astronomical Centre in Warsaw.  M.A.A. was supported
by the Swedish Research Council, the Polish grant N203 009 31/1466, and the 
Czech grant MSM 4781305903. E.S. was supported by the Czech grant LC 06014.



\begin{thebibliography}{}

\bibitem[2001]{AK01}
Abramowicz, M. A. \& Klu\'zniak, W. 2001,
A\&A, 374, L19

\bibitem[2004]{AK04}
Abramowicz, M. A. \& Klu\'zniak, W. 2004,
AIP Conference Proceedings, 714, 21A

\bibitem[2003]{abr03}
Abramowicz, M. A., Bulik, T., Bursa, M. \& Klu\'zniak, W. 2003,
A\&A, 404, L21

\bibitem[2005]{abr05}
Abramowicz, M. A. (editor) 2005,
"Nordita Workdays on QPOs", Astronomische Nachrichten, 326, 9

\bibitem[1978]{arn78}
Arnold, V. I. 1978, Mathematical methods of classical mechanics,
New York: Springer

\bibitem[1983]{arn83}
Arnold, V. I. 1983, Geometrical methods in the theory of ordinary 
differential equations, in
Grundlehren der Mathematischen Wissenschaften, New York: Springer

\bibitem[1991]{bal91}
Balbus, S. A. \& Hawley, J. F. 1991, 
ApJ, 376, 214

\bibitem[2005]{bel05}
Belloni, T., M{\'e}ndez, M. \& Homan, J. 2005,
A\&A, 437, 209

\bibitem[1985]{bla85}
Blaes, O. 1985,
MNRAS, 216, 553

\bibitem[2006]{bla06a}
Blaes, O. M., Arras, P. \& Fragile, C. P. 2006a,
MNRAS, 369, 1235

\bibitem[2006]{bla06b}
Blaes, O. M., \v Sr\'amkov\'a, E., Abramowicz, M. A., Klu\'zniak, W.
\& Torkelsson, U. 
2006b, ApJ, submitted

\bibitem[2004]{kat04}
Kato, S. 2004,
PASJ, 56, 905

\bibitem[2005]{kat05}
Kato, S. 2005,
PASJ, 57, 699

\bibitem[2000]{klis00}
van der Klis, M. 2000,
Ann. Rev. Astr. Ap., 38, 717

\bibitem[2000]{klu00}
Klu\'zniak, W. \& Abramowicz, M. A. 2000,
Phys. Rev. Lett. (submitted), [astro-ph/0105057]

\bibitem[2004]{klu04}
Klu\'zniak, W., Abramowicz, M. A. \& Lee, W. H. 2004,
ApJ, 603, 93L

\bibitem[2004]{klu042}
Klu\'zniak, W., Abramowicz, M. A., Kato, S., Lee, W. H., \&  Stergioulas, N. 2004,
ApJ, 603, L89

\bibitem[2004]{Lee04}
Lee, W. H., Abramowicz, M. A. \& Klu\'zniak, W. 2004,
ApJ, 603, L93

\bibitem[2004]{mc:rem}
McClintock, J. E., Remillard, R. A. 2006,
in Compact Stellar X-ray Sources, ed. W. H. G. Lewin \& M. van der Klis
(Cambridge: Cam. Univ. Press), 157-213, [astro-ph/0306213]

\bibitem[2004]{mon04}
Montero, P. J., Rezzolla, L. \& Yoshida, S. 2004,
MNRAS, 354, 1040

\bibitem[1980]{paczynski:wiita}
Paczy\'nski, B. \& Wiita, P. J. 1980,
A\&A, 88, 23

\bibitem[1995]{pap95}
Papaloizou, J. C. B. \& Lin, D. N. C. 1995,
ApJ, 438, 841

\bibitem[1983]{pap83}
Papaloizou, J. C. B. \& Pringle, J. E. 1983,
MNRAS, 202, 1181

\bibitem[1996]{pri96}
Pringle, J. E. 1996, 
MNRAS, 281, 357

\bibitem[2003]{rez03}
Rezzolla, L., Yoshida, S., Maccarone, T. J. \& Zanotti, O. 2003a,
MNRAS, 344, L37

\bibitem[2003]{rez032}
Rezzolla, L., Yoshida, S., Zanotti, O. 2003b,
MNRAS, 344, 978

\bibitem[2005]{rub05}
Rubio-Herrera, E. \& Lee, W. H. 2005a,
MNRAS, 357, L31

\bibitem[2005]{rub05}
Rubio-Herrera, E. \& Lee, W. H. 2005b,
MNRAS, 362, 789-798

\bibitem[1992]{sto92}
Stone, J. M. \& Norman, M. L. 1992, 
ApJS, 80, 791 

\bibitem[2001]{stro01}
Strohmayer, T. 2001,
ApJ, 554, L169

\bibitem[2006]{vio06}
Vio, R., Rebusco, P., Andreani, P., Madsen, H. \& Overgaard, R. V. 2006,
A\&A, 452, 383

\bibitem[2005]{zan05}
Zanotti, O., Font, J. A., Rezzolla, L. \& Montero, P. J. 2005,
MNRAS, 356, 1371


\end{thebibliography}
\end{document}